\documentclass[9pt,conference]{IEEEtran}

\usepackage[preprint]{waspaa25}
\usepackage{amsmath}

\usepackage{bm} 
\usepackage{multirow}
\usepackage[table]{xcolor} 
\usepackage{comment}
\usepackage{graphicx}      
\usepackage{adjustbox}
\usepackage{textcomp}
\usepackage{refcount}

\newcommand{\rpattern}{\mathbf{r}}
\newcommand{\rpatternElem}{r}
\newcommand{\rpIndex}{n}
\newcommand{\rpIndexNext}{{n'}}
\newcommand{\obsStrums}{\mathbf{s}}
\newcommand{\obsStrumsElem}{s}
\newcommand{\barIndex}{m}
\newcommand{\normalDist}{\mathcal{N}}
\newcommand{\likelihood}{L}
\newcommand{\timeSignature}{\mathcal{T}}

\newcommand{\costValue}{c}
\newcommand{\playedTiming}{t_\mathrm{p}}
\newcommand{\writtenTiming}{t_\mathrm{w}}
\newcommand{\dist}{d}

\newcommand{\repoURL}{\url{https://github.com/YousicianGit/rhythmic-pattern-transcription}}

\title{Transcribing Rhythmic Patterns of the Guitar Track in Polyphonic Music}

\name{Aleksandr Lukoianov,
      Anssi Klapuri\thanks{The financial support of Business Finland is gratefully acknowledged.}}
\address{Yousician, Helsinki, Finland}

\begin{document}

\maketitle

\begin{abstract}
Whereas chord transcription has received considerable attention during the past couple of decades, far less work has been devoted to transcribing and encoding the rhythmic patterns that occur in a song. The topic is especially relevant for instruments such as the rhythm guitar, which is typically played by strumming rhythmic patterns that repeat and vary over time. However, in many cases one cannot objectively define a single “right” rhythmic pattern for a given song section. To create a dataset with well-defined ground-truth labels, we asked expert musicians to transcribe the rhythmic patterns in 410 popular songs and record cover versions where the guitar tracks followed those transcriptions. 
To transcribe the strums and their corresponding rhythmic patterns, we propose a three-step framework. Firstly, we perform approximate stem separation to extract the guitar part from the polyphonic mixture. Secondly, we detect individual strums within the separated guitar audio, using a pre-trained foundation model (MERT) as a backbone. Finally, we carry out a pattern-decoding process in which the transcribed sequence of guitar strums is represented by patterns drawn from an expert-curated vocabulary. 
We show that it is possible to transcribe the rhythmic patterns of the guitar track in polyphonic music with quite high accuracy, producing a representation that is human-readable and includes automatically detected bar lines and time signature markers. We perform ablation studies and error analysis and propose a set of evaluation metrics to assess the accuracy and readability of the predicted rhythmic pattern sequence.
\end{abstract}

\section{Introduction}
\label{sec:intro}

Rhythmic information is complementary to the harmonic progression of a song.
However, transcription of rhythmic patterns has received much less attention than chord transcription, for example. Websites like \url{ultimate-guitar.com} often provide also strumming patterns of the songs, suggesting that those are valuable to the users of those sites -- often guitar players.
The most common way of playing the guitar 
is by strumming chords: 
that conveys the harmonic progression and the pulse (tempo and beat) intended.
By also choosing a rhythmic pattern that fits the song in question, the player can make the rhythmic feel of their performance more authentic.

\begin{figure}[t]
  \centering
  \centerline{\includegraphics[width=\columnwidth]{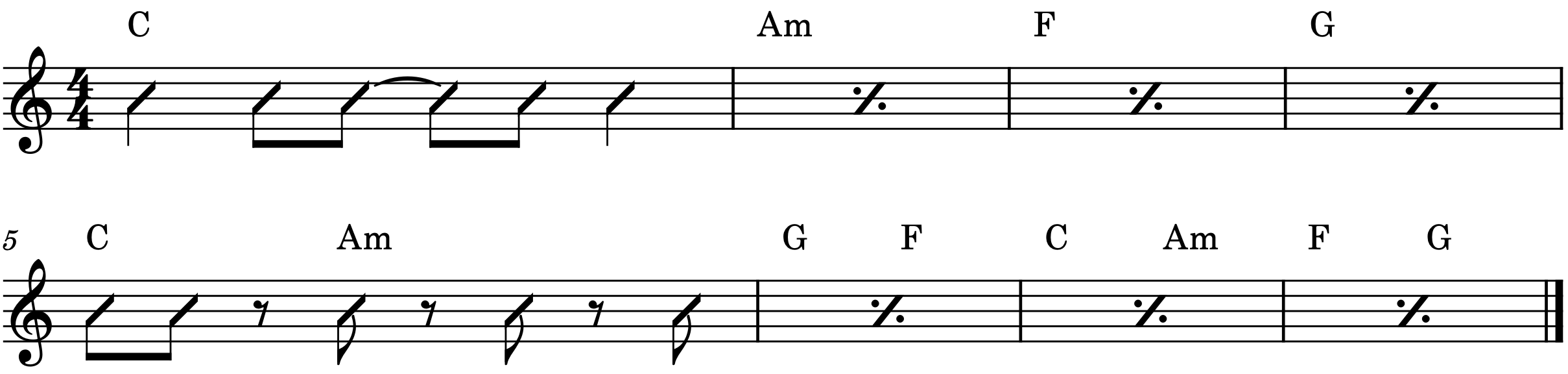}}
  \caption{Example rhythmic patterns written using the slash notation. The symbol $\%$ indicates that the written-out rhythm should be repeated.}
  \label{fig:slash_notation}
\end{figure}

\Cref{fig:slash_notation} shows an example of how rhythmic patterns can be notated.
In popular and jazz music, such
lead-sheet style ways of writing music are widely used. A guitar player often finds them more convenient to read 
than the full tablature, preferring to rely on their knowledge of the song and general musicianship when it comes to rendering the details of the performance. 

In this paper, we propose a method for transcribing the sequence of strums played by the rhythm guitar track in polyphonic music, and subsequently, represent the strum sequence with a sequence of rhythmic patterns drawn from a finite vocabulary defined by expert musicians. 
To create a dataset with well-defined ground truth labels, we asked expert musicians to transcribe the rhythmic patterns in 410 popular songs and record cover versions where the guitar track follows those transcriptions. Several versions were produced for each song, with difficulty levels for the guitar part ranging from a simplified version to the rhythms played by the original artists.
Using this data, we show that the considered task is practically feasible: meaningful rhythmic patterns of the guitar track can be extracted from polyphonic music and converted to a human-readable representation that includes automatically produced bar lines and time signature markers.

\section{Related work}
\label{sec:related_work}

Guitar strum detection is closely related to 
onset detection \cite{Bello2005, BöckDAFX2012, SchluterICASSP2014}. 
There has also been some work on guitar strum detection and classification specifically
\cite{Bello2019, Barbancho2014, Su2014, Murgul2022}, however, focusing on the analysis of isolated guitar tracks as opposed to polyphonic music.
There is a separate body of research addressing the classification of the overall rhythm of a polyphonic music, without focusing on any individual instrument \cite{DixonISMIR2004, Gouyon2004, Goiiyon2004, Tsunoo2009}.
This sometimes involves rhythmic similarity estimation and rhythmic pattern matching \cite{PaulusISMIR2002, Holzapfel2009, Guastavino2009, Panteli2014}.

The practical application considered here concerns lead-sheet level music transcription, such as that in \cref{fig:slash_notation}. 
Chord and tonality analysis is not covered in this paper,
but an excellent review can be found in \cite{PauwelsISMIR2019}. State of the art chord recognition systems typically apply deep neural networks (DNNs), including fully-connected \cite{Korzeniowski2016}, 
convolutional \cite{korzeniowski2016fully, humphrey2012rethinking} and recurrent DNNs \cite{sigtia2015audio, mcfee2017structured, hori2017music, Korzeniowski2018}. Most recently,
Transformers have been employed \cite{chen_harmony_2019, park2019bi}. 
As a part of the work in this paper, we perform bar line estimation, also called downbeat tracking.
We utilize the BeatThis method of Foscarin et al.
\cite{foscarin2024beatthis}, but other
state-of-the-art methods include \cite{DurandICASSP2015}, \cite{BöckISMIR2016} and \cite{HungICASSP2022}.

From a methodological viewpoint, large-scale pre-training is reshaping almost all MIR research.
Ma et al. \cite{ma2024foundation} review the trend, list main architectures and tuning schemes, and flag challenges such as long-context modeling and transfer evaluation.
Donahue et al.\cite{donahue2022melody}  demonstrate that Jukebox representations, combined with beat, key, and chord modules, enable direct lead-sheet transcription from audio.
Li et al. \cite{li2024mert} introduce MERT, a HuBERT-style encoder that adapts to beat, harmony and tagging.
Won et al. \cite{won2024foundation} show that Conformer can lead to superior performance when
fine-tuned on beat, chord, structure and tagging.
Pasini et al. \cite{music2latent} propose
Music2Latent; its frozen latents serve key, pitch-class and tagging tasks.
Hung et al. \cite{hung2023scaling} leverage 240 kh of music via noisy-student training, letting larger PerceiverTFs excel at downbeat, chord and structure.
Ding et al. \cite{ding2024parameter} find that adapters, LoRA and prompt tuning can match full fine-tuning for tagging, key and tempo while updating only a few weights.

\section{Data}
\label{sec:data}
The dataset 
that we use in this paper
comprises 931 proprietary recordings of 410 popular songs. Each song appears in up to four difficulty levels for the guitar part: simplified, intermediate, advanced, and original.
All versions preserve the core musical elements that make the song identifiable. The simplified, intermediate, and advanced variants may be shorter, transposed to an easier key, or feature streamlined rhythmic patterns, whereas the ``original'' version is arranged to match the album recording as closely as possible structurally, harmonically and rhythmically. Example tracks from the dataset are publicly available at GitHub.\footnote{\label{fn:repo}\repoURL}

For each song version, we have isolated backing, vocal, and guitar tracks, together with the corresponding transcription. The dedicated guitar track represents an acoustic steel-string guitar that plays the strumming patterns of interest. The backing tracks contain multiple instruments, often also other guitar(s), as the primary intention of the produced covers versions was to be faithful to the original song.

\begin{figure}[t]
  \centering
  \centerline{\includegraphics[width=\columnwidth]{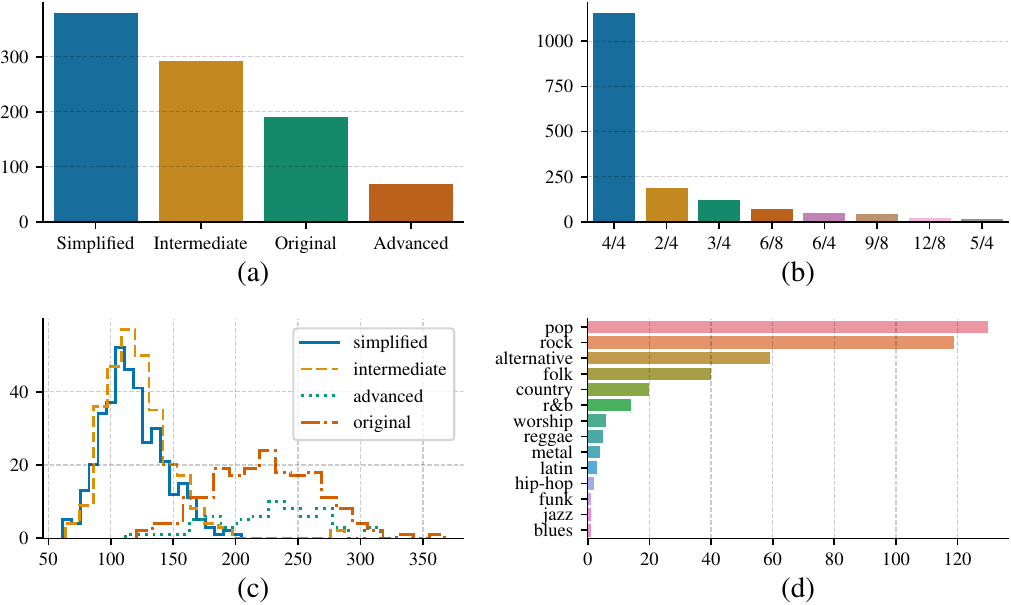}}
  \caption{Dataset overview: (a) difficulty levels, (b) time signatures, (c) track durations (seconds), and (d) genre distribution.
  }
  \label{fig:dataset_overview}
\end{figure}

\Cref{fig:dataset_overview} summarizes statistics of the dataset.
Most songs are in a 4/4 time signature, and the dominant genres are pop and rock.
Simplified and intermediate versions are noticeably shorter than the advanced and original versions.
In total, expert musicians identified 924 distinct rhythmic patterns in the dataset. Most of them span one or two bars, with the sixteenth note as their finest subdivision.

\section{Strum detection}
\label{sec:strum_detection}

To transcribe rhythmic patterns, we first need to detect individual strums in the polyphonic mixture.

\subsection{Stem separation}
\label{sec:stem_separation}

Isolated rhythm-guitar stems are rarely available in real-world scenarios, so we must operate on the full mix. Recent evaluations \cite{watcharasupat2024stem,fabbro2024} report good separation scores of roughly 10~dB SNR for vocals, bass, and drums, but barely above 0~dB SNR for the remaining instruments, including guitars. As separation artifacts can be quite detrimental to onset detection, we took the approach of suppressing the sources that can be removed with high quality (vocals, bass, and drums), producing a residual ``other'' stem. This signal is then fed to the downstream model that learns to pick out the guitar of interest implicitly. 

For source separation we chose the open-source HTDemucs \cite{rouard2023hybrid} model with four stems (vocals, drums, bass, other; we use ``other''), which performed best overall in our setup. We also experimented with the six-stem variant of HTDemucs (adding piano and guitar; we use guitar)
and a baseline that processes the full mix without any separation.

\subsection{Model}
\label{sec:method}

Because our dataset is relatively small, we rely on a foundation model pre-trained via self-supervised learning on large-scale, unlabeled music corpora and readily adaptable to MIR tasks.
We select MERT \cite{li2024mert} for its strong frame-level performance, particularly on beat tracking \cite{marble2023}, suggesting it captures temporal features useful for strum detection.

We investigate two training strategies: probing and fine-tuning. Probing keeps the encoder frozen while training only a lightweight task head. Because downstream tasks favor different encoder layers \cite{li2024mert}, the head receives a learnable, weighted average of all layer hidden states. Prior studies \cite{hung2023scaling,won2024foundation} show that fine-tuning often outperforms probing for beat tracking and especially for chord and key recognition. 
Our task -- detecting guitar strums in polyphonic mixtures -- may similarly benefit, as fine-tuning can encourage implicit separation of the guitar track in the encoder. 
Indeed, fine-tuning outperformed probing in all our experiments (\cref{tab:strum-detection-main-results}).

We largely follow the training and post-processing strategy introduced in BeatThis \cite{foscarin2024beatthis}.
During training we use the Shift-Tolerant Weighted Binary Cross-Entropy loss that ensures that small timing errors in the annotations are not penalized.
At inference we apply a simple post-processing step, peak-picking the frame with the highest probability above 0.5 inside every $\pm40$ms neighborhood. 
We found that these two ingredients 
are essential for strong performance.

The model combines a \texttt{MERT-v1-95M} encoder and an MLP with 256 units and 0.25 dropout that outputs a per-frame onset probability. 
In probing, the encoder is frozen.
For fine-tuning, all 12 encoder layers plus the positional convolution embedding are unfrozen.
We train
on full-length tracks with an effective batch of about 10 min of audio ($\approx 4$ songs). 
Following \cite{foscarin2024beatthis}, we select the checkpoint with the highest validation F1
-- even if the validation loss starts increasing -- to favor stable, high-confidence predictions and avoid spurious strums.

\subsection{Data augmentation}
\label{sec:data_augmentation}
Because the ground-truth stems feature only an acoustic steel-string guitar\cite{zang2024synthtab,pedrosa2024leveraging}, we synthesized additional guitar tracks to improve robustness to other timbres. For every song version, we generated one synthetic stem by randomly selecting one of the 80 in-house guitar presets that represent different recording and playing techniques of 11 acoustic and 6 electric guitars. The electric guitar tracks were further subjected to varying amounts of distortion effect. The synthetic stem was then mixed with the original backing track, effectively doubling the dataset.
We train our model on both original and synthetic audio and evaluate on three subsets: \emph{All} (played + synthetic), \emph{Played} (original recordings only), and \emph{Synth} (artificial mixes only).

At training time, each excerpt is independently transposed by choosing a random non-zero value from the range +2 to -3 semitones with 50\% probability. The shift is applied through resampling the time-domain audio signal, which affects the playback speed also, therefore, the annotation time-stamps were scaled by the same factor to stay aligned. This exposes the model to different keys and slight tempo variation while preserving the rhythmic structure.

\section{Extraction of rhythmic patterns}
\label{sec:patterns}
The idea of rhythmic pattern sequence decoding is to represent the observed (detected) strums with a sequence of rhythmic patterns drawn from an expert-curated vocabulary. Each pattern is either 1 or 2 musical measures long and starts and ends at a bar line. The sequence of patterns must cover the entire song from start to finish without gaps.
Each pattern has been further annotated with a time signature label, such as 4/4 or 6/8.
We add empty patterns in order to represent measures that do not contain any strums. 
Ten different empty patterns are added, one for each time signature.

At the level of an individual strum, we assume that the timing of each played strum, $\playedTiming$, is normally distributed around the nominal (written) timing: $\playedTiming \sim \normalDist(\writtenTiming, \sigma^2)$. The variance $\sigma^2$ models timing imperfections and is assumed to be constant throughout the song. 

The sequence of observed strums is obtained from the fine-tuned on ``other'' stem model (second-last row of \cref{tab:strum-detection-main-results}).

\begin{table*}[t]
\centering
\caption{Strum detection results by stem and training method. For every experiment, metrics are shown for the same stem the model was trained on.
Rows labeled with $^{\dagger}$ assume oracle access to isolated tracks; they represent an ideal upper bound and are therefore ignored when the best (max) scores are highlighted.
Standard errors for precision and recall closely match those of F1 and are omitted for brevity.}
\label{tab:strum-detection-main-results}

\sisetup{
    detect-weight,
    round-mode          = places,
    round-precision     = 1,
    separate-uncertainty,
    uncertainty-separator = {\,\pm\,},
    table-number-alignment = center,
}

\begin{adjustbox}{max width=\linewidth}
\begin{tabular}{ll*{9}{S[table-format=2.1]}}
  \toprule
  & & \multicolumn{3}{c}{\textbf{All}} 
    & \multicolumn{3}{c}{\textbf{{Played}}} 
    & \multicolumn{3}{c}{\textbf{Synth}}\\
  \cmidrule(lr){3-5}\cmidrule(lr){6-8}\cmidrule(lr){9-11}
  \textbf{Method} & \textbf{Stem} 
      & {F1} & {Precision} & {Recall}
      & {F1} & {Precision} & {Recall}
      & {F1} & {Precision} & {Recall}\\
  \midrule
  \multirow{1}{*}{Baseline}
      & isolated track$^{\dagger}$
        & \num{83.0 +- 0.3} & \num{79.7} & \num{89.1}
        & \num{81.4 +- 0.5} & \num{76.3} & \num{89.7}
        & \num{84.5 +- 0.4} & \num{83.1} & \num{88.6}\\
  \midrule
  \multirow{4}{*}{Probing (Frozen)}
      & isolated track$^{\dagger}$
        & \num{96.8 +- 0.1} & \num{96.3} & \num{97.5}
        & \num{94.7 +- 0.2} & \num{94.2} & \num{95.5}
        & \num{98.9 +- 0.1} & \num{98.5} & \num{99.5}\\
      & guitar stem
        & \num{87.0 +- 0.4} & \num{83.8} & \num{92.6}
        & \bfseries\num{87.5 +- 0.5} & \bfseries\num{84.8} & \bfseries\num{92.1}
        & \num{86.6 +- 0.6} & \num{82.9} & \num{93.2}\\
      & other stem
        & \bfseries\num{88.2 +- 0.3} & \bfseries\num{84.4} & \bfseries\num{94.4}
        & \num{86.7 +- 0.5} & \num{83.7} & \num{92.0}
        & \bfseries\num{89.6 +- 0.4} & \bfseries\num{85.2} & \bfseries\num{96.9}\\
      & full mix
        & \num{78.6 +- 0.4} & \num{72.0} & \num{91.4}
        & \num{77.0 +- 0.6} & \num{70.9} & \num{89.2}
        & \num{80.1 +- 0.5} & \num{73.2} & \num{93.6}\\
  \midrule
  \multirow{4}{*}{Fine-tuning}
      & isolated track$^{\dagger}$
        & \num{98.3 +- 0.1} & \num{98.4} & \num{98.3}
        & \num{96.7 +- 0.2} & \num{96.8} & \num{96.7}
        & \num{99.9 +- 0.0} & \num{99.9} & \num{99.9}\\
      & guitar stem
        & \num{95.4 +- 0.2} & \num{95.2} & \num{96.2}
        & \num{94.8 +- 0.3} & \num{94.8} & \num{95.3}
        & \num{96.1 +- 0.3} & \num{95.6} & \num{97.1}\\
      & other stem
        & \bfseries\num{96.9 +- 0.2} & \bfseries\num{96.7} & \bfseries\num{97.6}
        & \bfseries\num{95.3 +- 0.3} & \bfseries\num{95.1} & \num{96.0}
        & \bfseries\num{98.5 +- 0.2} & \bfseries\num{98.2} & \bfseries\num{99.1}\\
      & full mix
        & \num{96.5 +- 0.2} & \num{96.0} & \num{97.5}
        & \num{94.9 +- 0.3} & \num{94.4} & \bfseries\num{96.1}
        & \num{98.1 +- 0.1} & \num{97.6} & \num{98.9}\\
  \bottomrule
\end{tabular}
\end{adjustbox}
\end{table*}

\subsection{Pattern sequence decoding}
\label{sec:decoding}
The pattern sequence decoding task can be viewed as a search problem: finding a sequence of patterns that minimizes the error in representing the observed strum sequence. 
We chose to tackle this problem with dynamic programming,  
as the amount of data we have was deemed insufficient for training an end-to-end neural network for the task. 
More specifically, we use the Viterbi algorithm to find the optimal pattern sequence for each song \cite{forney2005}. 

Viterbi algorithm requires two quantities be defined: \textit{observation probabilities} $p(\obsStrums_\barIndex | \rpattern_\rpIndex)$ that define the probability of observing the strum sequence $\obsStrums_\barIndex$ in measure $\barIndex$ given a candidate rhythmic pattern $\rpattern_\rpIndex$; and \textit{transition probabilities} $P(\rpattern_\rpIndexNext(t) | \rpattern_\rpIndex(t - 1))$ that define the probability that pattern $\rpattern_\rpIndex$ is followed by pattern $\rpattern_\rpIndexNext$.

For calculating the observation probabilities, we employ a variant of the two-way mismatch method \cite{maher1994}. The probability that the observed (detected) strum sequence $\obsStrums_\barIndex$ in measure $\barIndex$ was generated by rhythmic pattern candidate $\rpattern_\rpIndex$ is calculated as
\begin{equation}\label{eq:two_way_mismatch}
   p(\obsStrums_\barIndex | \rpattern_\rpIndex) =
   \prod_i \normalDist 
   \left( \dist(\obsStrumsElem_{\barIndex,i} | \rpattern_\rpIndex); 0, \sigma^2 \right)
   \prod_j \normalDist 
   \left( \dist(\rpatternElem_{\rpIndex,j} | \obsStrums_\barIndex); 0, \sigma^2 \right)
\end{equation}
where $\obsStrumsElem_{\barIndex,i}$ denotes the i:th element of $\obsStrums_\barIndex$ and
$
    \dist(\obsStrumsElem_{\barIndex,i} | \rpattern_\rpIndex) =
        \min_j | \obsStrumsElem_{\barIndex,i} - \rpatternElem_{\rpIndex, j}|
$
is the distance of $\obsStrumsElem_{\barIndex,i}$ to the nearest strum within the candidate pattern $\rpattern_\rpIndex$. Symmetrically, $\rpatternElem_{\rpIndex,j}$ denotes the j:th element in the candidate rhythmic pattern, and $\dist(\rpatternElem_{\rpIndex,j} | \obsStrums_\barIndex)$ denotes its distance to the nearest element among the observed strums $\obsStrums_\barIndex$.

The values within the two vectors, $\obsStrums_\barIndex$ and $\rpattern_\rpIndex$ are between zero and one, expressing the position of the strum within a single measure.
If a rhythmic pattern is two-measures long, the first measure is matched against $\obsStrums_\barIndex$ and the second measure is matched against $\obsStrums_{\barIndex + 1}$.

In practice, the optimization process employs unnormalized log-likelihood values $\likelihood(\cdot)$ and omits any normalizing constants, simplifying \cref{eq:two_way_mismatch} to 
$   
   \likelihood(\obsStrums_\barIndex | \rpattern_\rpIndex) \propto
   \sum_i \dist(\obsStrumsElem_{\barIndex,i} | \rpattern_\rpIndex)^2 +
   \sum_j \dist(\rpatternElem_{\rpIndex,j} | \obsStrums_\barIndex)^2 + C
$,
where the additive constant $C$ can be dropped too. 
As a special case, if both $\rpattern_\rpIndex$ and $\obsStrums_\barIndex$ are empty, we set $\likelihood(\rpattern_\rpIndex | \obsStrums_\barIndex)$ to zero. If only $\rpattern_\rpIndex$ or $\obsStrums_\barIndex$ is empty, we set the likelihood to $-\infty$.

The other quantity, transition probabilities, play to role of favouring pattern continuity and thereby \textit{readability} of the resulting transcription. That consists of two elements: preferring to repeat the same rhythmic pattern where possible, and secondly, preferring transitions between patterns that have been labeled with the same time signature.

We define the (unnormalized) transition log-probabilities as:
\begin{equation*}\label{eq:transition_probs}
   L(\rpattern_\rpIndexNext(t) | \rpattern_\rpIndex(t - 1)) = \begin{cases}
0 & \text{if $\rpIndexNext = \rpIndex$}\\
-\costValue_1 &\text{if $\rpIndexNext \neq \rpIndex$ and $\timeSignature(\rpattern_\rpIndexNext) = \timeSignature(\rpattern_\rpIndex)$} \\
-\costValue_1 -\costValue_2  &\text{if $\rpIndexNext \neq \rpIndex$ and $\timeSignature(\rpattern_\rpIndexNext) \neq \timeSignature(\rpattern_\rpIndex)$} \\
\end{cases} 
\end{equation*}
where $\timeSignature(\rpattern_\rpIndex)$ denotes the time signature of pattern $\rpattern_\rpIndex$ and the constants $\costValue_1$ and $\costValue_2$ were found experimentally.

We use a flat prior distribution $p(\rpattern_\rpIndex)$ for the patterns, as favoring patterns that were more common in the training data was not helpful.

\subsection{Bar line estimation}
\label{sec:barlines}

Bar line estimation, also called downbeat detection, is an indispensable part of rhythm transcription because bar lines play a big role for human readability.
We chose the BeatThis method of Foscarin et al. due to its very good performance in downbeat detection and its ability to handle different time signatures
\cite{foscarin2024beatthis}.
A characteristic of the method is that it does not employ a dynamic Bayesian network for post-processing.
As a result, the model output sometimes exhibits discontinuities: the lengths of musical measures may suddenly double or halve, or the placing of bar lines may slip into the middle of a measure. 

We implemented a post-processing method for BeatThis that enforces consistent bar line placing 
without sacrificing the performance of the method too much.
The post-processing is based on a steady-tempo assumption\footnote{This assumption does not hold for our dataset: the amount of songs with drastic tempo changes is representative of the genres involved, and the described post-processing usually makes things worse for this minority of songs.}
and uses dynamic programming to lock into a temporally stable bar line sequence that best matches the unprocessed estimates from BeatThis. That is achieved by allowing individual bar lines to be deleted, or musical measures to be subdivided by small integer factors.
Full description is beyond the scope of this paper, therefore we publish our Python implementation as open source.\footnotemark[\getrefnumber{fn:repo}] 

\begin{table*}[t]
\centering
\caption{Rhythmic pattern sequence extraction results. \textit{Full system} refers
to the preceding row where both the transition probabilities were included.}
\label{tab:pattern_decoding_results}
\sisetup{
  detect-weight,
  round-mode      = places,
  round-precision = 1,
  separate-uncertainty,
  uncertainty-separator = {\,\pm\,},
  table-number-alignment = center
}
\begin{tabular}{ll*{8}{S[table-format=2.1]}}
  \toprule
    & & \multicolumn{3}{c}{\textbf{Reconstructed strum sequence $\uparrow$}} 
    & \multicolumn{3}{c}{\textbf{Discontinuity rate (\%) $\downarrow$}} \\
  \cmidrule(lr){3-5}\cmidrule(lr){6-8}
  \textbf{Method} & \textbf{Test data} & {F1} & {Precision} & {Recall}
    & {Patterns}
    & {Time signatures} & {Measure lengths} \\
  \midrule
  Ground truth bar lines & {All levels} &
    \num{96.8 +- 0.2} & \num{96.5} & \num{97.6} & 
    \num{24.0 +- 0.5} & \num{13.1 +- 0.3} & \num{0.90 +- 0.07} \\ 
  \midrule
  BeatThis bar lines & {All levels} &
    \bfseries\num{95.3 +- 0.2} & \num{95.0} & \num{96.0} & 
    \num{30.0 +- 0.5} & \num{15.3 +- 0.3} & \num{4.88 +- 0.2} \\
  BeatThis with post-proc. & {All levels} &
    \num{94.7 +- 0.2} & \num{94.3} & \num{95.5} & 
    \num{27.8 +- 0.5} & \num{14.6 +- 0.3} & \bfseries\num{0.41 +- 0.02} \\
  $+$ pattern transition cost & {All levels} &
    \num{94.7 +- 0.2} & \num{94.5} & \num{95.6} & 
    \bfseries\num{15.4 +- 0.3} & \num{9.4 +- 0.2} & \num{0.41 +- 0.02} \\
  $+$ time sign. trans. cost & {All levels} & 
    \num{94.7 +- 0.2} & \num{94.5} & \num{95.6} & 
    \num{15.4 +- 0.3} & \bfseries\num{0.1 +- 0.02} & \num{0.41 +- 0.02} \\
  \midrule
  \multirow{4}{*}{Full system} & Simplified &
    \num{94.3 +- 0.4} & \num{93.6} & \num{95.7} & 
    \num{11.9 +- 0.4} & \num{0.1 +- 0.02} & \num{0.46 +- 0.04} \\    
   & Intermediate &
    \num{95.6 +- 0.3} & \num{95.7} & \num{96.0} & 
    \num{17.2 +- 0.6} & \num{0.2 +- 0.04} & \num{0.46 +- 0.04} \\
   & Advanced &
    \num{94.4 +- 0.9} & \num{93.6} & \num{95.6} & 
    \num{18.6 +- 1} & \num{0.1 +- 0.05} & \num{0.22 +- 0.04} \\
   & Original &
    \num{94.5 +- 0.5} & \num{94.6} & \num{94.7} & 
    \num{18.3 +- 0.8} & \num{0.2 +- 0.05} & \num{0.25 +- 0.03} \\
  \bottomrule
\end{tabular}
\end{table*}

\section{Results}
\label{sec:results}
We employ 5-fold cross-validation. For each fold, the data are divided into training, validation, and test sets, ensuring that all difficulty levels  of a given song remain in the same partition. The five test folds are disjoint and together cover the entire dataset. We compute metrics per song version and report their mean and standard error of the mean.

\subsection{Strum detection}
\label{sec:strum_results}

As a baseline, strum onsets are detected by peak picking in the onset-strength envelope using \texttt{librosa} package \cite{mcfee2015librosa}. The peak-picking hyperparameters are tuned on the combined training and validation sets for 100 trials. We report this baseline only for the isolated guitar track, as it provides an approximate upper bound for the other stems. Onset-level precision, recall, and F1 are computed with \texttt{mir\_eval} \cite{raffel2014mir_eval} using a 50ms tolerance.

In \cref{tab:strum-detection-main-results} we explore training methods and stem separation effect. Every configuration beats the baseline except probing on the full mixtures, underscoring the need for (approximate) stem separation when frozen MERT is used as a feature extractor. Fine-tuning always outperforms probing, and the gap widens as the input becomes closer to the full mixture, indicating that updating encoder layers helps the model focus on the target guitar in polyphonic audio. The best scores achieved with the “other” stem, yet a fine-tuned full-mix model attains nearly the same performance.

Metrics on \emph{Synth} are consistently higher than those on \emph{Played}. The primary cause is the expressive variability of the human performances in the \emph{Played} set. Expert musicians do not always follow the transcription \textit{precisely}, altering rhythmic patterns or introducing subtle onset misalignments (humans cannot hit every beat exactly). By contrast, \emph{Synth} tracks are artificially generated \textit{directly} from the score and therefore contain no such annotation noise. This also explains almost 100\% metrics for fine-tuned isolated-track model. 

Unexpectedly, the evaluation metrics for the guitar stem are lower than those for the “other” stem. 
Manual inspection points to two main causes. First, as discussed in \cref{sec:stem_separation}, the quality of the guitar-stem separation is generally lower.
Second, the backing tracks often include additional guitar parts (see \cref{sec:data}) that play different rhythms or techniques such as strumming or arpeggiated chords. These extra parts are sometimes amplified in the guitar stem and divert the model’s attention from the guitar track of interest.

\subsection{Bar line estimation}
\label{sec:barline_resuts}

\Cref{tab:barline_results} shows results for bar line estimation, compared against ground truth bar lines annotated by expert musicians for our dataset.
The F1, precision and recall metrics were computed with the \texttt{mir\_eval} toolbox, 
using a 70ms tolerance to be consistent with \cite{foscarin2024beatthis}.
Discontinuity rate is calculated by counting the number of measures where the length differs more than 35\% from the length of the previous measure, and then dividing the count by the total number of measures. 

\begin{table}[t!]
\centering
\caption{Bar line estimation results.}
\label{tab:barline_results}
\sisetup{
  detect-weight,
  round-mode      = places,
  round-precision = 1,
  separate-uncertainty,
  uncertainty-separator = {\,\pm\,},
  table-number-alignment = center
}
\begin{adjustbox}{max width=\linewidth}
\begin{tabular}{l*{4}{S[table-format=2.1]}}
  \toprule
  & \textbf{F1} & \textbf{Precision} & \textbf{Recall} & \textbf{Discont. (\%)} 
  \\
  \midrule
  Ground truth &
         \num{100.0 } & \num{100.0} & \num{100.0} & \num{0.90 +- 0.07} \\
  BeatThis downbeats &
        \num{88.4 +- 0.3} & \num{88.7} & \num{90.5} & \num{4.88 +- 0.2} \\        
  $+$ post-processing &
    \num{87.2 +- 0.4} & \num{87.5} & \num{90.2} & \bfseries\num{0.41 +- 0.02} \\
  \bottomrule
\end{tabular}
\end{adjustbox}
\end{table}

As can be seen from the table, the ground-truth bar line annotations have slightly less than 1\% of discontinuities. Raw BeatThis output has 4.88\%. While that number is quite low, it still affects the human readability quite a lot.
After our proposed post-processing, the amount of discontinuities collapses to 0.41\% -- at the cost of decreasing bar line detection F1 measure from 88.4\% to 87.2\%.

\subsection{Pattern sequence decoding}
\label{sec:pattern_resuts}

\Cref{tab:pattern_decoding_results} shows the results for the pattern sequence decoding. 
The quality of the produced transcriptions is assessed from two main viewpoints: 1) accuracy and 
2) human readability of the produced transcription.
The most intuitive way that we found to measure the accuracy of the transcription is to reconstruct the strum sequence by ``writing out'' the produced rhythmic pattern sequence,
and then compare the reconstructed strum sequence with the ground truth. 
using a 50ms tolerance to allow direct comparison with the results in \cref{tab:strum-detection-main-results}.

In order to evaluate the human readability of the produced transcription, we found it most efficient to monitor the rate of discontinuities in the produced patterns, time signatures, and measure lengths. Pattern discontinuity is calculated by counting the number of times the rhythmic pattern changes during a song, and dividing that by the total number of measures.
For example value 24.0\% in the table means that each rhythmic pattern continues for $1.0/0.24 \approx 4.2$ bars on the average. 
Time signature discontinuity rate is calculated by counting the number of time signature changes during the song and dividing that by the number of measures within the song.

When using ground-truth bar lines, the F1 metric is 96.8\% -- almost same as the F1 metric for the estimated strums \textit{before} representing them with a sequence of patterns (96.9\%, see \cref{tab:strum-detection-main-results}). 
When using estimated bar lines from BeatThis, the F1 measure of the reconstructed strum sequence drops to 95.3\%. When further enforcing continuity of the estimated bar lines with our proposed post-processing technique, the F1 measure drops slightly further to 94.7\%, however also reducing the amount of bar line discontinuities from 4.88\% to 0.41\%.

Adding non-flat transition probabilities does not decrease the accuracy of the transcription at all.
However, readability increases clearly, as pattern change rate drops from 27.8\% to 15.4\%
and the time signature discontinuity rate even more drastically, from 9.4\% to 0.1\%.
Notably, the proposed system achieves time signature estimation as a by-product, since the transition probabilities force the decoder to stick to a consistent time signature most of the time.

The last four rows of \cref{tab:pattern_decoding_results} show pattern sequence decoding results for the different difficulty categories. Here ``full system'' refers to the last row in the previous section of the table, where both the transition probabilities were included.
Most notable is that the amount of pattern discontinuities is lower in the \textit{simplified} arrangements than in the \textit{original} or other arrangements (11.9\% vs. 18.3\%).

\section{Conclusions}
\label{sec:conclusions}

We have described techniques for transcribing the sequence of strums played on the guitar track in polyphonic music. The strum sequence was then further translated into a sequence of rhythmic patterns drawn from an expert-curated vocabulary. The resulting transcription included automatically estimated bar lines and time signature markers.
The results indicate that the considered task is practically feasible and can lead to an accurate and human-readable transcription. We therefore encourage others also to consider this task in their future work.

\clearpage
\bibliographystyle{IEEEtran}
\bibliography{refs25}
\end{document}